\documentclass{ceab}   

\usepackage{epsfig}     
\usepackage{graphicx}   

\usepackage{sidecap} 

\usepackage{ceabbib}     
\usepackage[bookmarks = true,colorlinks = true,linkcolor = blue,urlcolor = blue,citecolor = blue,bookmarks = true,linktocpage = true,hyperindex = true,breaklinks]{hyperref}

\usepackage[T1]{fontenc}

\begin{document}

\title{Empirical atmosphere model in a mini flare during magnetic reconnection}

\author{Brigitte Schmieder$^{1,2}$, Reetika Joshi$^3$, 
Ramesh Chandra$^{3}$, \\
Guillaume Aulanier$^4$,
Akiko Tei$^5$, Petr Heinzel$^6$, James Tomin$^1$,\\
Nicole Vilmer$^1$, and  V\'eronique Bommier$^1$
\vspace{2mm}\\
\it $^1$LESIA, Observatoire de Paris, Université PSL, CNRS, Sorbonne Université,\\
\it  Université de Paris, 5 place Jules Janssen, 92190 Meudon, France\\ 
\it $^2$Centre for mathematical Plasma Astrophysics, Dept. of Mathematics,\\
\it KU Leuven, 3001 Leuven, Belgium\\ 
\it $^3$ Department of Physics, DSB Campus, Kumaun University,\\ 
\it Nainital-263001, India\\
\it $^4$  Laboratoire de Physique des Plasmas (LPP), École Polytechnique,\\
\it IP Paris, Sorbonne Université, CNRS, Observatoire de Paris,\\ 
\it Université PSL, Université Paris Saclay, Paris, France\\
\it $^5$ Institute of Space and Astronautical Science, Japan, Aerospace Exploration Agency,\\
\it 3-1-1 Yoshinodai, Chuo-ku, Sagamihara,
Kanagawa 252-5210, Japan\\
\it $^6$  Astronomical Institute of the Czech Academy of Sciences, Fričova 298, 251 65\\
\it Ond\v{r}ejov, Czech Republic\\
}
\def\gore{Schmieder, B., et al.}
\maketitle

\begin{abstract}
A spatio-temporal analysis of IRIS spectra of Mg\textsc{ii}, C\textsc{ii}, and Si\textsc{iv} ions allows us to study the dynamics and the stratification of the flare atmosphere along the line of sight during the magnetic reconnection phase at the jet base. Strong asymmetric Mg\textsc{ii} and C\textsc{ii} line profiles with extended blue wings observed at the reconnection site are interpreted by the presence of two chromospheric temperature clouds: one explosive cloud with blueshifts at 290 km s$^{-1}$ and one cloud with smaller Doppler shift (around 36 km s$^{-1}$). Simultaneously at the same location a mini flare was observed with strong emission in multi temperatures (AIA), in several spectral IRIS lines (e.g. O\textsc{iv} and Si\textsc{iv}, Mg\textsc{ii}), absorption of identified chromospheric lines in Si\textsc{iv} line profile, enhancement of the Balmer continuum and X-ray emission by FERMI/GBM. With the standard thick-target flare model we calculate the energy of non thermal electrons observed by FERMI and compare it to the energy radiated by the Balmer continuum emission. We show that the low energy input by non thermal electrons above 20 keV was still sufficient to produce the excess of Balmer continuum.
\end{abstract}


\section{Introduction}

Solar flare models have been extensively developed using 1D  to  3D non LTE radiative transfer codes  in flare loops \citep{Allred2020,Kerr2020}.  The sites of chromospheric excitation during solar flares are marked by  white light emission, extended extreme ultraviolet ribbons and hard X-ray footpoints. The standard interpretation is that these are the result of heating and bremsstrahlung emission  from non-thermal electrons precipitating from the corona \citep{Fletcher2013}. 
Solar flare models  interpret   enhancements of emission in white light and in Balmer continuum by non thermal electron beams \citep{Heinzel2014,Kleint2016,Kleint2017,Kowalski2017}.
It is important to understand the relationship between electron beams and radiative emission  for interpreting   white light super flares in stars  \citep{Kowalski2012,Heinzel2018}.

We report on a solar jet observed  in multi wavelength with  the \emph{Interface Region Imaging Spectrograph} \citep[IRIS,] [] {Pontieu2014} (IRIS) in chromospheric lines and with  the {\it Atmospheric Imaging Assembly} \citep[AIA,] [] {Lemen2012} on board  the \emph{Solar Dynamics Observatory} \citep[SDO,] [] {Pesnell2012} occurring on March 22 2019.
This event has been well described  using spectroscopy  data  of IRIS and spectropolarimetry data  from HMI   in a series of papers \citep{Joshi2020FR,Joshi2021a,Joshi2021b} and    NLFF  extrapolation  of the photosphere magnetic field  have been performed  
\citep{Yang2020}.

Here we present a summary of the observations leading to an empirical  atmosphere model  of the mini flare at  the base of the jet (UV burst).
The mini flare (GOES B6.7) occurred  in the active region  (AR NOAA 12736), which was located at N09 W60 on March 22, 2019. 
\begin{figure*}[ht!]
\centering
\includegraphics[width=0.75\textwidth]{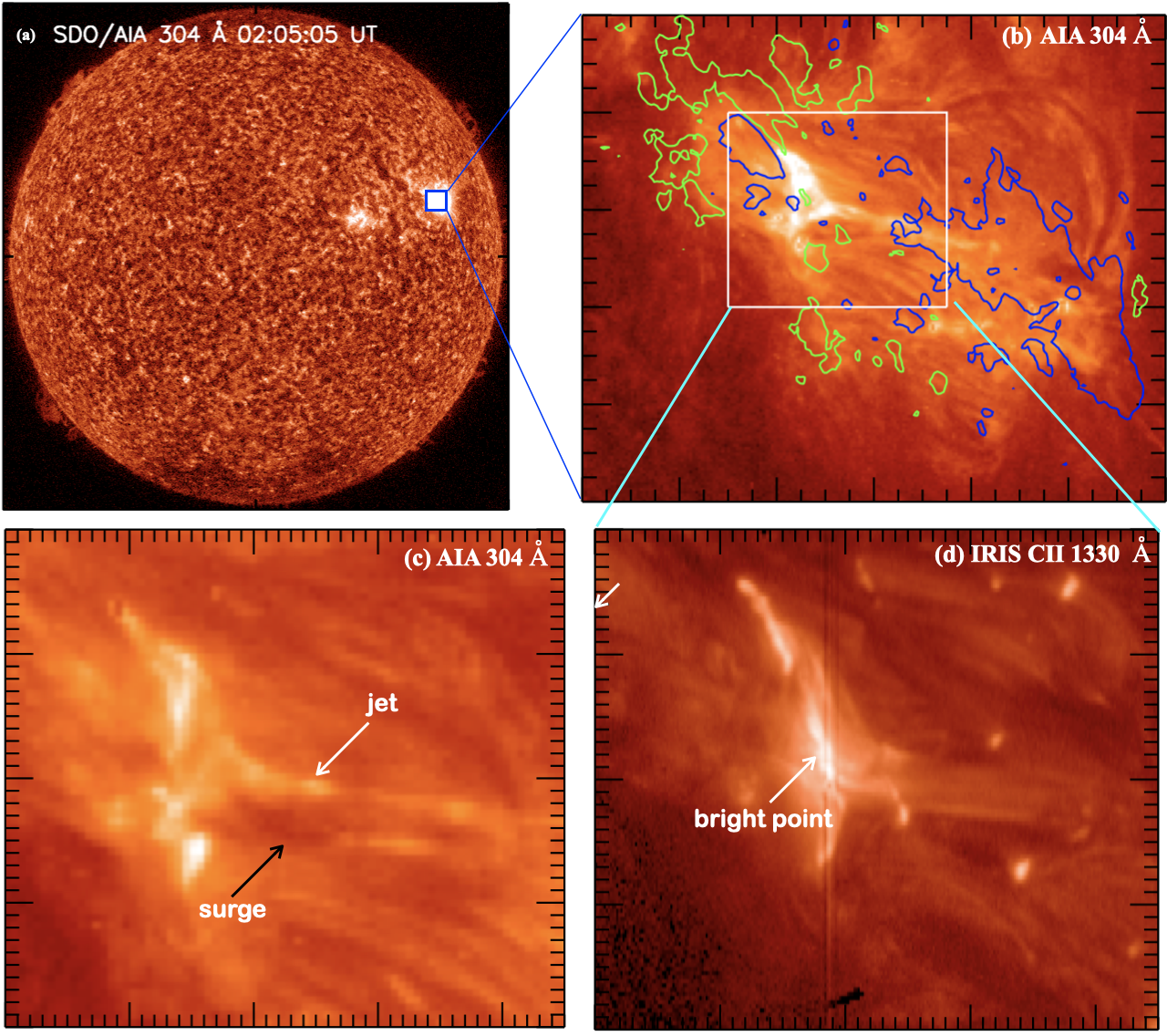}
\caption{Observations of the solar jet and surge on March 22, 2019. Panel a shows the full disk image of the Sun, and the blue rectangular box is the AR zoomed in panel b. In panel b the green and blue contours represent positive and negative magnetic polarity respectively ($\pm$ 300 Gauss). Magnetic reconnection occurred between two emerging magnetic flux regions. The jet and cool surge are indicated in panel c. The reconnection site (bright point) is crossed by the IRIS slit position 1 indicated in panel d.}
\label{fulldisk}
\end{figure*}

\begin{figure*}[t!]
\centering
\includegraphics[width=0.8\textwidth]{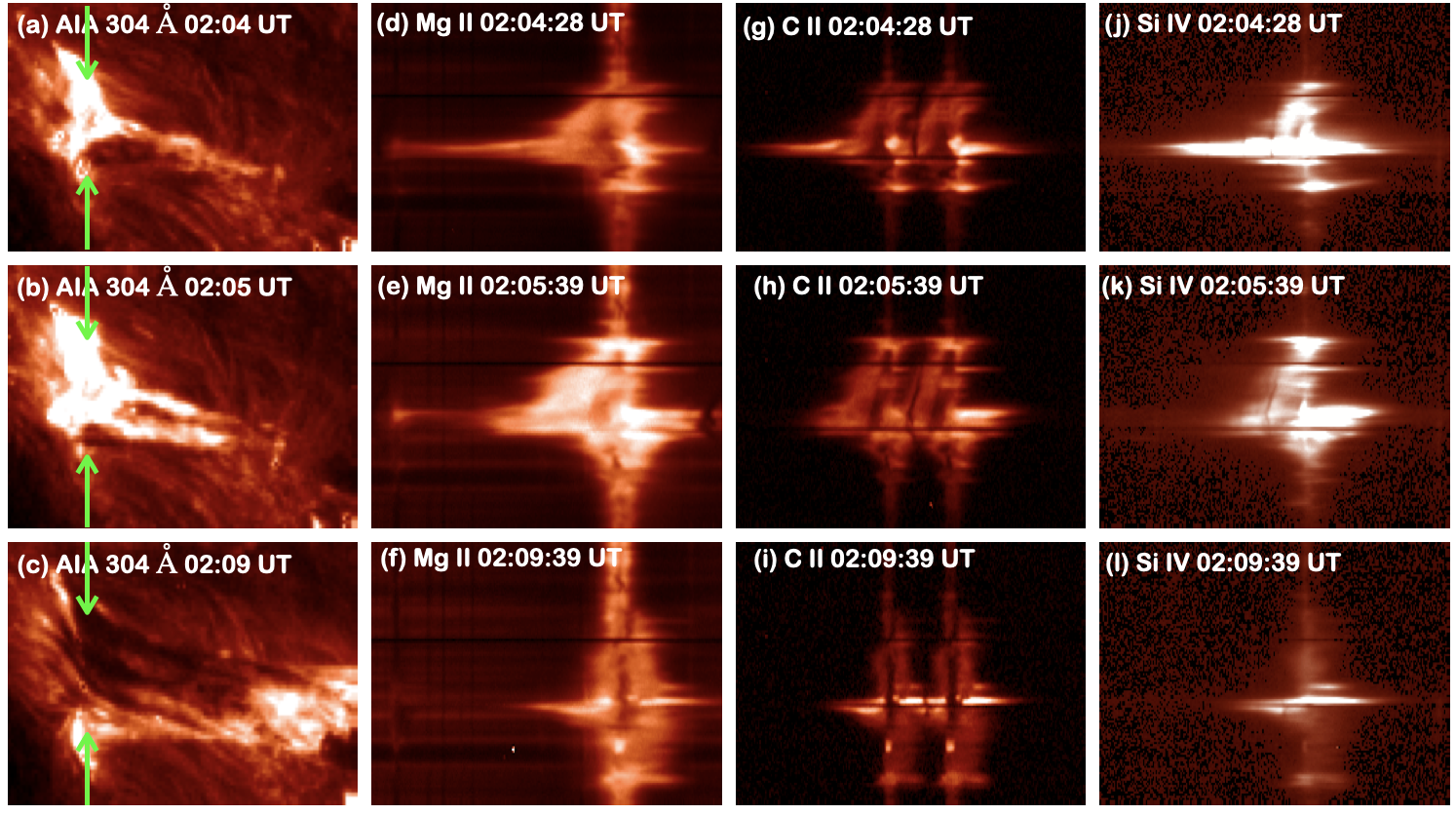}
\caption{Jet reconnection base (UV burst/mini-flare) and jet evolution. First column: images in AIA 304 \AA. Second, third, and last columns show IRIS spectra of the jet reconnection site at the slit position 1 (shown by green arrows).}
\label{IRIS}
\end{figure*}

\begin{SCfigure}
\centering
\includegraphics[width=0.4\textwidth]{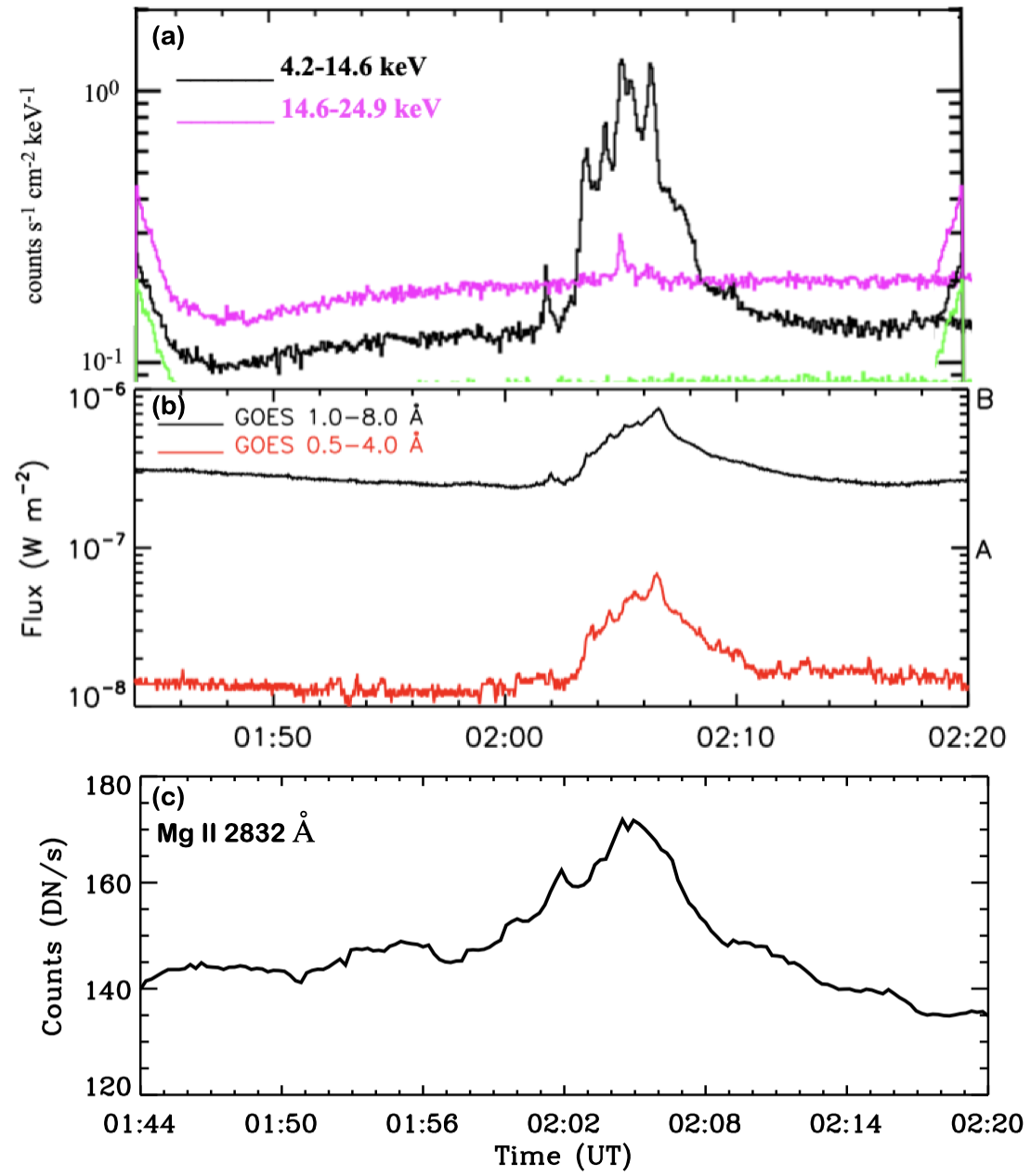}
\caption{Hard X-ray emission recorded by FERMI/GBM  (panel a), GOES soft X-ray  (panel b), light curve of the intensity  in the bright point in the Balmer continuum  image  obtained from the  IRIS SJI  at 2832 \AA\ (panel c).}
\label{goes}
\end{SCfigure}

\section{Observations}
The AR was the only AR in the whole solar disk on that day (Fig.  \ref{fulldisk} panel a).
The AR   was formed during the day and the previous day, by   successive emerging fluxes.
At the time of the jet at 02:04 UT  two emerging flux collapse and a flux rope was created between them \citep{Joshi2020FR}. Figure  \ref{fulldisk}  panel (b) shows an AIA 304 image with overlying magnetic field contours which indicate the presence of
positive and negative polarities  on both side of the brightening. These polarities  are squeezed and  at the location of the bright arch center   a small bipole (blue and green) is detected. The reconnection occurs at this point and a  bright vault is observed  from which escapes the jet. In panel (c) the bright jet and a parallel  dark surge  are observed.  The bright point is indicated in panel (d).  This bright point is observed in all the channels of AIA indicating  that the plasma is heated to more than 1 MK and corresponds to the B 6.7 X-class detected by GOES.

IRIS spectra have been analysed  showing the evolution of the jet and the mini flare at its base in the three band passes  centered in  Mg\textsc{ii}, C\textsc{ii} and Si\textsc{iv}  lines (Fig. \ref{IRIS}).  The spectra along the slit shows in all the panels at 02:04:28 UT an extended horizontal brightening, indicating large flows and  heating. These flows are characterized  as bidirectional flows during the reconnection. The extension of the flows have been calculated with the cloud model technique and  two clouds were identified with flows of -36 km s$^{-1}$ and  -300 km s$^{-1}$. The brightening is also  extended far away of Mg II k line blue wing indicating an enhancement of the Balmer continuum.   The Balmer continuum presents an excess  of brightness by 50\%.

We combine the observations of   Balmer continuum obtained with  IRIS  (spectra and SJI 2832 \AA) and   hard X-ray emission detected by  FERMI  Gamma Burst Monitor (GBM) during  the reconnection producing the  mini flare (Fig. \ref{goes}). Calibrated Balmer continuum is compared to non-LTE radiative transfer flare  models  and  
 radiated energy is estimated.  With the standard thick-target flare model we calculate the energy of non-thermal electrons  detected by  FERMI GBM  and compare it to  the radiated energy \citep{Joshi2021b}. We need to assume a very small area for the reconnection site, probably smaller then the IRIS pixel resolution for a satisfactory fit.
 
\section{Conclusion}
\begin{SCfigure}
\centering
\includegraphics[width=0.6\textwidth]{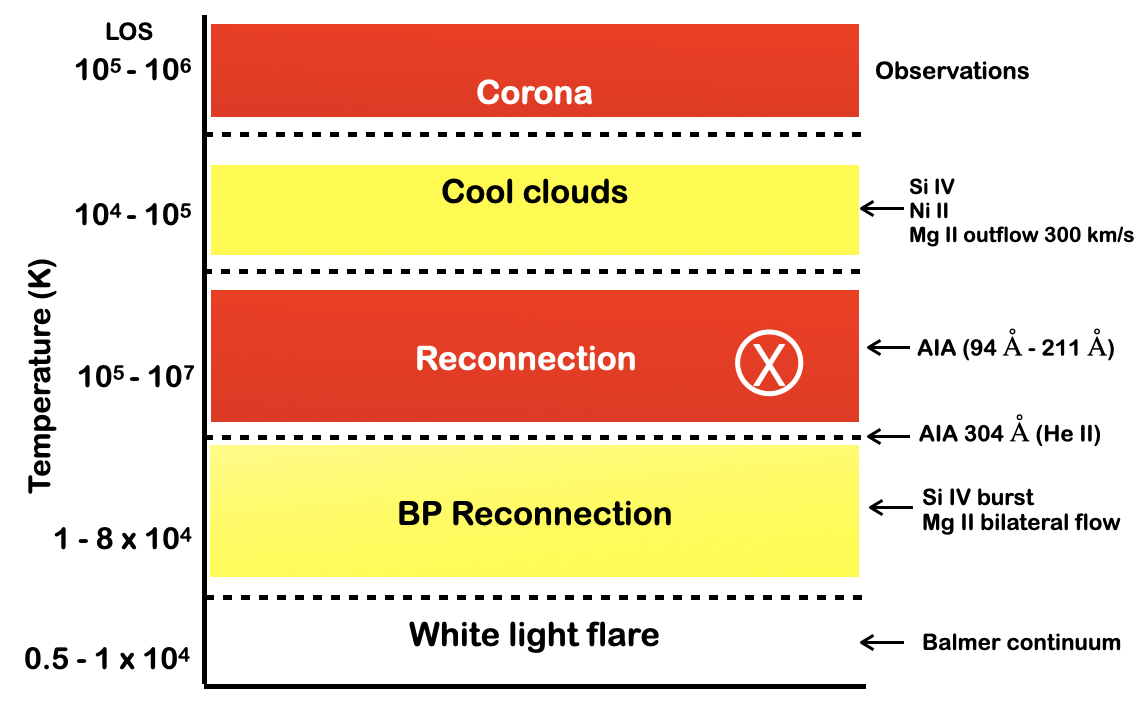}
\caption{Empirical flare model along the line of sight (LOS) through the current sheet observed at the reconnection point deduced from the analysis of the IRIS spectra and  AIA images.}
\label{model}
\end{SCfigure}

From the study of the IRIS spectra and filter images we deduce  an empirical flare model  for the mini flare (UV burst) at the base of the jet during reconnection (Fig. \ref{model}).  A sandwich of different temperature layers  explain the observations. It  is consistent with MHD models \citep{Hansteen2019}.

\bibliographystyle{ceab}
\bibliography{refs}

\end{document}